\documentclass[sigconf]{acmart}
\usepackage{enumitem}
\usepackage{colortbl}
\usepackage{enumerate }
\usepackage{subfigure}
\usepackage{amsmath, amsthm}
\AtBeginDocument{%
  }

\setcopyright{acmlicensed}
\copyrightyear{2025}
\acmYear{2025}
\setcopyright{cc}
\setcctype{by}
\acmConference[MM '25]{Proceedings of the 33rd ACM International Conference on Multimedia}{October 27--31, 2025}{Dublin, Ireland}
\acmBooktitle{Proceedings of the 33rd ACM International Conference on Multimedia (MM '25), October 27--31, 2025, Dublin, Ireland}\acmDOI{10.1145/3746027.3755779}
\acmISBN{979-8-4007-2035-2/2025/10}
\begin{document}

%%
%% The "title" command has an optional parameter,
%% allowing the author to define a "short title" to be used in page headers.
\title{Refining Contrastive Learning and Homography Relations for Multi-Modal Recommendation}

%%
%% The "author" command and its associated commands are used to define
%% the authors and their affiliations.
%% Of note is the shared affiliation of the first two authors, and the
%% "authornote" and "authornotemark" commands
%% used to denote shared contribution to the research.
\author{Shouxing Ma}
\affiliation{%
  \institution{University of Technology Sydney}
  \city{Sydney}
  \country{Australia}
}
\email{shouxingmaa@gmail.com}
\orcid{0009-0000-4964-4323}

\author{Yawen Zeng}
\affiliation{%
  \institution{Hunan University}
  \city{Changsha}
  \country{China}
}
\email{yawenzeng11@gmail.com}
\orcid{0000-0003-1908-1157}

\author{Shiqing Wu}
\authornotemark[1]
\affiliation{%
    \department{Faculty of Data Science}
  \institution{City University of Macau}
  \city{Macau SAR}
  \country{China}}
\email{sqwu@cityu.edu.mo}
\orcid{0000-0001-6785-1203}

\author{Guandong Xu}
\authornote{Corresponding Authors.}
\affiliation{%
  \institution{The Education University of Hong Kong}
  \city{Hong Kong SAR}
  \country{China}
}
\email{gdxu@eduhk.hk}
\orcid{0000-0003-4493-6663}

%%
%% By default, the full list of authors will be used in the page
%% headers. Often, this list is too long, and will overlap
%% other information printed in the page headers. This command allows
%% the author to define a more concise list
%% of authors' names for this purpose.
% \renewcommand{\shortauthors}{Trovato et al.}

%%
%% The abstract is a short summary of the work to be presented in the
%% article.
\begin{abstract}
 Multi-modal recommender system focuses on utilizing rich modal information ( i.e., images and textual descriptions) of items to improve recommendation performance. The current methods have achieved remarkable success with the powerful structure modeling capability of graph neural networks. However, these methods are often hindered by sparse data in real-world scenarios. Although contrastive learning and homography ( i.e., homogeneous graphs) are employed to address the data sparsity challenge, existing methods still suffer two main limitations: 1) Simple multi-modal feature contrasts fail to produce effective representations, causing noisy modal-shared features and loss of valuable information in modal-unique features; 2) The lack of exploration of the homograph relations between user interests and item co-occurrence results in incomplete mining of user-item interplay.

To address the above limitations, we propose a novel framework for \textbf{R}\textbf{E}fining multi-mod\textbf{A}l cont\textbf{R}astive learning and ho\textbf{M}ography relations  (\textbf{REARM}). Specifically, we complement multi-modal contrastive learning by employing meta-network and orthogonal constraint strategies, which filter out noise in modal-shared features and retain recommendation-relevant information in modal-unique features. To mine homogeneous relationships effectively, we integrate a newly constructed user interest graph and an item co-occurrence graph with the existing user co-occurrence and item semantic graphs for graph learning. The extensive experiments on three real-world datasets demonstrate the superiority of REARM to various state-of-the-art baselines. Our visualization further shows an improvement made by REARM in distinguishing between modal-shared and modal-unique features. Code is available \href{https://github.com/MrShouxingMa/REARM}{here}.
\end{abstract}

%%
%% The code below is generated by the tool at http://dl.acm.org/ccs.cfm.
%% Please copy and paste the code instead of the example below.
%%

%%
%% Keywords. The author(s) should pick words that accurately describe
%% the work being presented. Separate the keywords with commas.
%%
%% This command processes the author and affiliation and title
%% information and builds the first part of the formatted document.
\maketitle

\begin{figure}[t]
	\centering
	\includegraphics[width=\columnwidth]{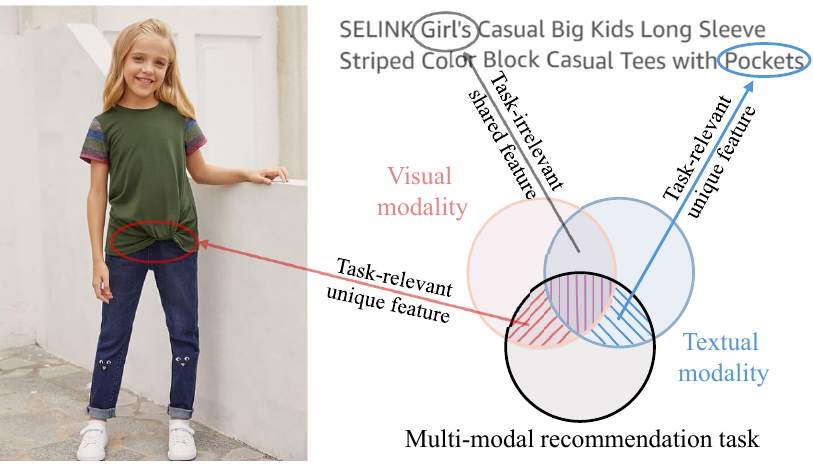}
 \caption{Illustration of contrastive learning for modalities in the multi-modal recommendation. Obtaining modal-shared features only by contrastive learning ignores valuable modal-unique features, and risks modal-shared features containing noise (irrelevant for the multi-modal recommendation task).}
\label{fig: intro}
\Description{...}
\end{figure}

\section{Introduction}
Recommender systems have become indispensable tools in contemporary e-commerce for discovering items of interest to users based on their preferences and behaviors \cite{smith2017two, covington2016deep}. Technological advances make it easier for users to access a wealth of multi-modal information about items, such as images, texts, and videos. By exploiting these rich contents, multi-modal recommender systems could obtain more accurate inferences about user interests or preferences compared to general recommender systems \cite{zhou2023bootstrap, wei2019mmgcn, zhang2021mining}.

Early researchers integrate multi-modal information into the classical collaborative filtering (CF) framework by either concatenating or summing multi-modal information \cite{he2016vbpr, liu2017deepstyle}. Recent approaches \cite{guo2024lgmrec, zhou2023tale} consider the graph structures for user-item interactions and multi-modal information, and explore potential higher-order connectivity with the help of graph neural networks (GNNs). Current state-of-the-art multi-modal recommender systems require high-quality supervised data to achieve optimal performance. However, observed interactions are extremely sparse in real e-commerce scenarios, compared to the entire interaction space~\cite{he2016ups}. This sparsity hinders the model's ability to learn efficiently, thereby limiting the performance of recommender systems \cite{wu2021self, lin2022improving,li2024federated}.

Inspired by the success of Self-Supervised Learning (SSL), researchers address data sparsity by creating self-supervised signals with multi-modal information \cite{zhang2022latent, yu2023multi}. Unlike in computer vision~\cite{yu2023multi} and natural language processing \cite{lan2019albert}—where SSL trains models for downstream tasks—most multi-modal recommendations~\cite{zhou2023bootstrap, guo2024lgmrec} construct self-supervised signals by contrasting features for the final representation. For instance, DiffMM \cite{jiang2024diffmm} employs cross-modal contrastive learning to align multi-modal information and reduce noise, improving user preference learning. While exploring SSL, other efforts \cite{zhang2021mining, wang2021dualgnn, zhang2022latent} explore multi-modal information and historical interactions to create homogeneous graphs, to alleviate the issue of sparse data. For example, DRAGON~\cite{zhou2023enhancing} combines dual representations from both heterogeneous and homogeneous graphs (i.e., user co-occurrence graph \& item semantic graph). Although current studies show promising results in multi-modal recommendations, they still face two significant limitations: \textbf{1) Simple contrasting multi-modal features leads to incomplete user and item representations.} Directly fusing multi-modal features fails to take into account unique and shared information is insufficient~\cite{song2024quest,liang2024factorized,dufumier2024align,wang2022rethinking}. On the one hand, unique features may be lost when aligning for consistency with contrastive learning~
\cite{tian2020makes, liang2021multibench}. For the example illustrated in Figure \ref{fig: intro}, the model can well maintain the consistency features through multi-modal contrastive learning, i.e., modal-shared features, in the overlapping regions of visual and textual modalities. However, there are modal-unique features that are simultaneously relevant to the multi-modal recommendation task. For instance, a fashionable twist design might be evident in the image but not in the text, or ``pockets'' mentioned in the text might not be clear from the image alone in Figure \ref{fig: intro}. Missing discriminative modal-unique information leads to suboptimal recommendation performance. Preserving valuable modal-unique features when exploring modal consistency is essential for achieving high-quality recommendations.
On the other hand, not all modal-shared features extracted by the model are useful; some may be noisy, irrelevant, or even misleading~\cite{wang2022rethinking,liang2024factorized}. For example, an item in Figure \ref{fig: intro} is identified as a girl's shirt by both image and text, but it could also be for boys based on the ``Big Kids'' tag. Insufficient and noisy feature representations may substantially degrade the performance of the model. Hence, the ability to retain salient modal-unique features
and eliminate modal-shared noise is vital for the effectiveness of multi-modal self-supervised models.
\textbf{2) Neglect of user interest and item co-occurrence relationships.} A substantial amount of informative signals are embedded within user interest patterns and item co-occurrence relationships, which can be leveraged to facilitate a deeper understanding of user behaviors and item semantics~\cite{liang2016factorization,chen2022cocnn,zhang2023beyond,liu2024attribute}. However, existing methods primarily focus on user co-occurrence and item semantic relationships while overlooking the associations between user interests and item co-occurrence patterns~\cite{zhou2023enhancing,zhang2022latent,zhou2023tale}. Hence, it is crucial to incorporate both co-occurrence and semantic (interest) relationships of users and items to enhance model representation capabilities.

\begin{figure*}[t]
% \vspace{-10pt}
	\centering
	\includegraphics[width=1\textwidth]{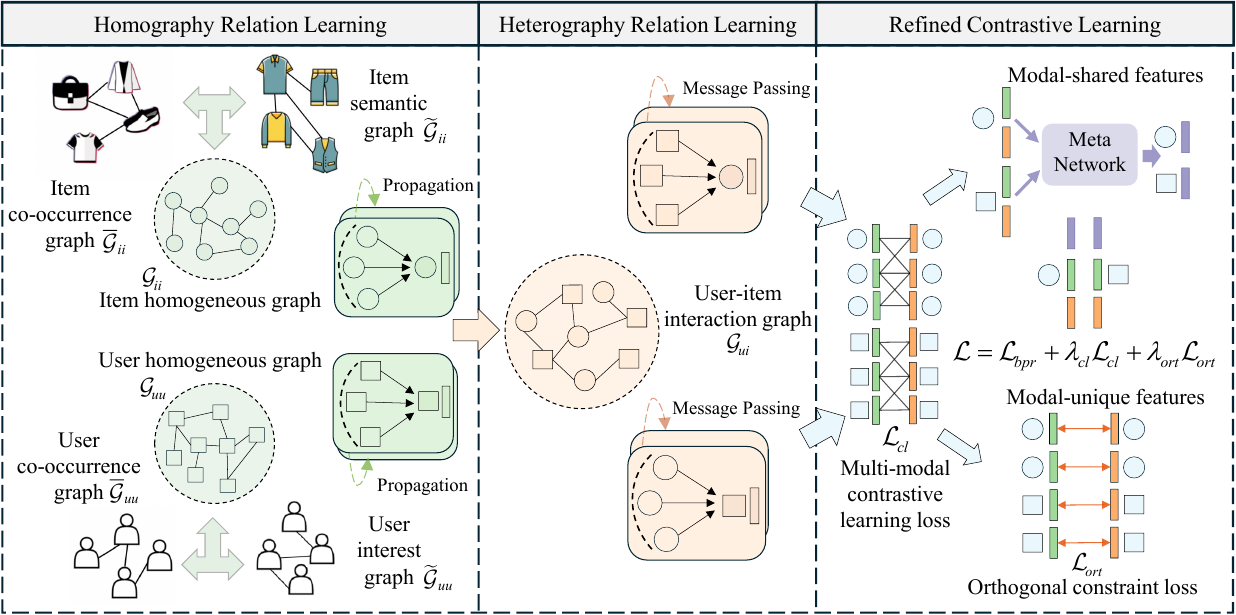}
	\caption{The structure overview of the proposed REARM.} 
\label{fig: model}
% \vspace{-10pt}
\Description{...}
\end{figure*}

To address the above issues, we propose a novel framework for \textbf{R}\textbf{E}fining multi-mod\textbf{A}l cont\textbf{R}astive learning and latent ho\textbf{M}ography relations  (\textbf{REARM}). Our overall approach (Figure \ref{fig: model}) comprises three core components: homography relation learning, heterography relation learning, and refining contrastive learning. First, we utilize the existing interactions and multi-modal information to additionally construct an item co-occurrence graph and user interest graph based on previous studies \cite{zhou2023enhancing,  yu2023multi}. The item semantic graph and the user co-occurrence graph, together, form user and item homogeneous graphs, respectively. These graphs enable the model to explore potential relationships among users and items from different perspectives of structure and semantics (interest). Next, we explore potential higher-order interactions from the interaction graph for each modality independently. We follow previous studies~ \cite{jiang2024diffmm, guo2024lgmrec} to introduce an auxiliary task to alleviate the data sparsity, i.e., we contrast multi-modal features to generate self-supervised signals. To further filter out noises in modal-shared features and preserve the modal-unique and recommendation-relevant information after contrasting, we refine the multi-modal contrastive learning by introducing the meta-network and orthogonal constraint techniques. The former leverages customized transformation matrices to extract recommendation-relevant information from modal-shared features to filter out noise, while the latter utilizes the orthogonal constraint loss function to encourage multi-modal features to retain modal-unique information. 
We validate the effectiveness of our framework on three public datasets, and our experimental results demonstrate distinct advantages of our model. Furthermore, we visualize the difference in the probability of interaction before and after refining the contrasting multi-modal features, which clearly shows the superiority of our method in distinguishing between modal-shared features and modal-unique features.

Our main contributions are summarized as follows.
\begin{itemize}[leftmargin=*]
\setlength{\itemsep}{2pt}
\setlength{\parsep}{2pt}
\setlength{\parskip}{2pt}
\item We propose a novel multi-modal contrastive recommendation framework (REARM), which preserves recommendation-relevant modal-shared and valuable modal-unique information through meta-network and orthogonal constraint strategies, respectively.

\item We jointly incorporate co-occurrence and similarity graphs of users and items, allowing more effective capturing of the underlying structural patterns and semantic (interest) relationships, thereby enhancing recommendation performance.

\item Extensive experiments are conducted on three publicly available datasets to evaluate our proposed method. The experimental results show that our proposed framework outperforms several state-of-the-art recommendation baselines.
\end{itemize}

\section{Related Work}
\subsection{Multi-modal Recommendation}
Due to the sparse interaction of recommender systems in the real world, numerous studies~\cite{wei2019mmgcn, wang2021dualgnn} have leveraged multi-modal data about users or items (e.g., images, descriptions, or reviews). Earlier work, such as VBPR \cite{he2016vbpr}, directly splices visual features from items into the vanilla CF framework. Later, GNNs shown to be capable of exploring potential higher-order interactions are introduced and largely utilized~\cite{wei2019mmgcn, yu2023multi}. Specifically, LGMRec~\cite{guo2024lgmrec} combines local graph and global hypergraph embedding modules to enhance representations.
Recently, SSL has received massive attention for the benefits of non-addition of data~\cite{xv2024improving}. For instance, DiffMM~\cite{jiang2024diffmm} improves representations by contrasting multi-modal features.

\subsection{GNN-based Recommendation}
The powerful higher-order connectivity of GNNs further extends the expressive power of CF models \cite{wang2019neural}. Along this line, LightGCN~\cite{he2020lightgcn} further investigates the importance of each component in GNNs, and experimentally removes feature transformation and nonlinear activation modules to better fit the recommendation. Later on, GNN-based methods are widely adopted as an essential component of various recommendations \cite{ chen2023heterogeneous, wei2019mmgcn}. For example, DRAGON~\cite{zhou2023enhancing} proposes a dual graph learning representation, which incorporates the user co-occurrence graph and the item semantic graph.

\subsection{Contrastive Learning for Recommendation}
Contrastive learning has been widely practiced in various fields ~\cite{chen2020simple, lan2019albert}, since it benefits from generating self-supervised signals based on the original data. This advantage also renders it a powerful tool in tackling the inherent data sparsity issue in recommender systems~\cite{lin2022improving, chen2023heterogeneous}. More specifically, SLMRec \cite{tao2022self} performs dropout and masking operations on multi-modal features to construct additional self-supervised signals. Also, similar multi-modal contrastive learning strategies are employed in DiffMM~\cite{jiang2024diffmm} to capture and utilize the modality-related consistency information.

\section{Theoretical Guarantee}
\subsection{Multi-modal Contrastive Learning}
Let $X_1$ and $X_2$ denote the different data modalities, while $Z$ and $Y$ denote the latent variable and the task, respectively. Multi-modal contrastive learning as a universal framework yields a latent variable $Z$ by maximizing the mutual information $I(X_1, X_2)$ to satisfy task $Y$ \cite{liang2024factorized}. However, this strategy is only effective under the assumption of multi-view redundancy \cite{wang2022rethinking, dufumier2024align, liang2021multibench}.

\textbf{Definition 1 (Multi-view redundancy).} $\exists \varepsilon > 0,$ such that
$I(X_1; Y|X_2) \leq \varepsilon$  and $I(X_2; Y| X_1) \leq \varepsilon.$

This assumption indicates that most of the task-relevant information is shared and that the unique information is as small as $\varepsilon$. While multi-view redundancy is correct for particular types of multi-modal data, valuing unique information as well as ignoring redundant irrelevant information is also critical \cite{liang2021multibench,wang2022rethinking, song2024quest}. Furthermore, Liang et al.~\cite{liang2024factorized} factorize task-relevant information to shared and unique information with separate optimizations. The same applies to multi-modal recommendations, and simple contrasting multi-modal features leads to incomplete representations. Hence, considering multi-modal contrastive learning, we further filter out recommendation-irrelevant information in modal-shared and retain recommendation-relevant information in modal-unique.

\subsection{Orthogonal Constraint}
A matrix, \( \mathbf{W} \), is orthogonal if \( \mathbf{W}^\top \mathbf{W} = \mathbf{W} \mathbf{W}^\top = \mathbf{I} \). Enforcing an orthogonality constraint between matrices could penalize and reduce the redundant information between them \cite{wang2010semi,salzmann2010factorized,wang2015deep}. Moreover, soft orthogonal constraint loss is defined to approximate orthogonality efficiently \cite{bousmalis2016domain,hazarika2020misa,liu2017adversarial}. Similar to previous work~\cite{ruder2018strong,zheng2024mulan}, we utilize orthogonal constraint loss to encourage non-redundancy retention between modals, and further incorporate classical recommendation loss to preserve valuable modal-unique information.

\section{Methodology}

\subsection{Preliminaries}
We conceptualize the user-item interactions as a bipartite graph $\mathcal{G}_{ui} = \{(u, i, e)|u\in \mathcal{U}, i\in \mathcal{I}, e\in \mathcal{E}\}$, where $\mathcal{U}$ denotes the set of $N$ users, $\mathcal{I}$ denotes the set of $M$ items, and $\mathcal{E}$ denotes the set of edges representing the observed interactions. Furthermore, we denote the multi-modal information derived for each item in pre-trained models as $m \in \mathcal{M}$. The modality feature of the item $i$ is represented as $\mathbf{i}_{m} \in \mathbb{R}^{d_{m}}$, where $d_{m}$ denotes the feature dimension. The modal features (interest preferences) of the user $u$ can be easily calculated as the mean of the modal features of all items observed interacting with him/her, i.e., $\mathbf{u_m}=\frac{1}{\left | \mathcal{I}_u \right | }  {\textstyle \sum_{i\in\mathcal{I}_u}^{}} \mathbf{i}_m$, where $\mathcal{I}_u$ denotes the set of all items that the observed the user $u$ has interacted with. In this work, we consider only two mainstream modalities, visual modality $v$ and textual modality $t$, i.e., $\mathcal{M} = \{v, t\}$. Nevertheless, our model could be easily extended to scenarios with more than two modalities. The goal of this work is to predict unobserved interactions between users and items given multi-modal and interaction information.

\subsection{Homography Relation Learning}
Prior work~\cite{yu2023multi, xv2024improving, zhang2022latent} has shown that mining potential inter-item associations could further enrich item representations. Moreover, other research~\cite{wang2021dualgnn, zhou2023enhancing} verifies the effectiveness of leveraging user-side co-occurrence for performance gains. However, potential associations between user interests and item co-occurrence patterns remain under-explored. Hence, we jointly explore co-occurrence and similarity relations from both user and item perspectives. 

For clarity, we illustrate the process using item examples, as user and item relationships are built similarly. Notably, homography is constructed before training, thus avoiding additional training costs. Moreover, data sparsity further alleviates computational overhead.

\subsubsection{Item Co-occurrence Graph Construction} Similar to establishing user co-occurrence relationships \cite{zhou2023enhancing}, we construct the item co-occurrence graph $\bar{\mathcal{G}}_{ii}= \{ \mathcal{I}, \bar{\mathcal{C}}_{ii} \}$, where $\bar{\mathcal{C}}_{ii} = \{ \bar{e}_{i, {i}' }|i, {i}' \in \mathcal{I}\}$ denotes the set of edges $\bar{e}_{i, {i}'}$ between item $i$ and item ${i}'$ in $\bar{\mathcal{G}}_{ii}$. We reserve the items with top-k common interactions among items and take the number as the weight value, which can be denoted as
\begin{align}
\bar{e}_{i, {i}' } =
\begin{cases}
\bar{e}_{i, {i}' }, & \text{if ${i}' \in top\mbox{-}k(i),$ }\\
0, & \text{otherwise.}
\end{cases}
\end{align}
Furthermore, as with \cite{wang2021dualgnn} handling co-occurrence, the softmax function is adopted to differentiate among item contributions.

\subsubsection{Item Semantic Graph Construction} Analogous to studies \cite{zhang2022latent, zhou2023enhancing}, we construct the modality-aware item semantic graph $\tilde{\mathcal{G}}^m_{ii} = \{ \mathcal{I}, \tilde{\mathcal{C}}^m_{ii} \}$ for each modality $m$, where $\tilde{\mathcal{C}}_{m}^{i}= \{\tilde{e}_{i, {i}' }^{m}| i, {i}' \in \mathcal{I}\}$ denotes the set of edges between item nodes in $\tilde{\mathcal{G}}_m$. And we calculate the similarity $s_{i{i}'}$ between the two items ($i$ and ${i}'$) with cosine similarity, and take it as the value of weight between them. Following the experience of~\cite{zhang2022latent}, we also perform the sparsification~\cite{chen2009fast} of the item semantic graph, and retain only the edges with high similarity. Specifically, the edge weights of the items could be formulated as
\begin{align}
\tilde{e}_{i, {i}' }^{m} =
\begin{cases}
s_{i, {i}' }^{m}, & \text{if $ s_{i, {i}' }^{m} \in top\mbox{-}k  ( s_i^{m})$}, \\
0, & \text{otherwise.}
\end{cases}
\end{align}
To mitigate potential gradient explosion or vanishing \cite{kipf2016semi},  degree normalization is applied as in~\cite{zhang2022latent}. Then, all modal semantic graphs are fused by summation using modal importance coefficients $\alpha ^{item}_m$, which sum up to 1.  Finally, we obtain the item semantic graph $\tilde{\mathcal{G}}_{ii} = \{ \mathcal{I}, \tilde{\mathcal{C}}_{ii} \}$, where $\tilde{\mathcal{C}}_{ii}= \{\tilde{e}_{i, {i}' }| i, {i}' \in \mathcal{I}\}$, $\tilde{e}_{i, {i}' }= {\textstyle \sum_{m \in \mathcal{M}}\alpha^{item}_{m}\tilde{e}_{i, {i}' }^{m}}$.

\subsubsection{Homogeneous Relations Learning} Given differences between the item co-occurrence graph $\bar{\mathcal{G}}_{ii}$ and the item semantic graph $\tilde{\mathcal{G}}_{ii}$ in various scenarios, we introduce the item co-occurrence factor $\alpha ^{item}_{co}$ to distinguish the final contribution, $\mathcal{G}_{ii}=\alpha ^{item}_{co}\bar{\mathcal{G}}_{ii}+(1-\alpha ^{item}_{co})\tilde{\mathcal{G}}_{ii}$.

Before exploiting higher-order associations by performing message passing on graph $\mathcal{G}_{ii}$, we first convert the item modality features $\mathbf{i}^{raw}_m$ to align with the ID embedding dimension. Formally,
\begin{align}
\mathbf{i}_m=\mathbf{W}^i_m\mathbf{i}^{raw}_m+\mathbf{b}^i_m,
\label{convert}
\end{align}
where $\mathbf{W}^i_m \in \mathbb{R}^{d\times d_{m}}$ and $\mathbf{b}^i_m \in \mathbb{R}^{d\times 1}$ are the transformation and bias parameters, and $d$ is the ID embedding dimension. With the consideration that the item IDs contain information that is distinct from the multi-modal features, we concatenate them to obtain the item representation $\mathbf{h}_i$, and perform homography relation learning,
\begin{align}
\mathbf{h}_{i}^{( l+1 ) } =\sum_{i^{'} \in \mathcal{N}_{i} }
e_{i, {i}' }\mathbf{h}_{i^{'}}^{(l)},
\label{ii_agg}
\end{align}
where $\mathcal{N}_{i}$ denotes the set of neighbors of item $i$ in the item homogeneous graph $\mathcal{G}_{ii}$. $\mathbf{h}_i^{(0)}$ is initialized with spliced IDs and multi-modal features, $\mathbf{h}_i^{(0)}=\mathbf{i}_{id}||\mathbf{i}_v||\mathbf{i}_t$. And we take the final layer output as the item homography representation. The item ID $\mathbf{i}_{id}^{hom}$ and multi-modal features $\mathbf{i}_m^{hom}$ learned through homography relations are richer in semantic and co-occurrence information.

Similarly, we derive the user co-occurrence graph $\bar{\mathcal{G}}_{uu}$, the interest graph $\tilde{\mathcal{G}}_{uu}$, and the homogeneous graph $\mathcal{G}_{uu}$ in sequence. Following Equation (\ref{ii_agg}), the operation on $\mathcal{G}_{uu}$ yields the user representations ($\mathbf{u}_{id}^{hom}$ and $\mathbf{u}_{m}^{hom}$) carrying more user's internal relations.

\subsubsection{Item Feature Attention Integration} 
Since the multi-modal features are not tailored for the recommendation task \cite{zhong2024mirror, xv2024improving} and GNNs may further amplify the noise in the modality \cite{liu2023multimodal}, we further refine by introducing self-attention and cross-attention modules.

The self-attention module better adapts to the downstream recommendation task by adaptively adjusting the magnitudes of the values of the dimensions within the modality. Formally,
\begin{align}
\mathbf{i}_m^{h-self}=softmax(\frac{ (\mathbf{i}_{m}^{hom}\mathbf{W}^{Q}_{m} )^{\top }(\mathbf{i}_{m}^{hom}\mathbf{W}^{K}_{m} )}
{\sqrt{d} } )\mathbf{i}_{m}^{hom}\mathbf{W}^{V}_{m},
\end{align}
where  $\mathbf{W}^{Q}_{m}, \mathbf{W}^{K}_{m}, \mathbf{W}^{V}_{m}  \in \mathbb{R}^{d\times d}$ are  parameter matrices, $softmax(\cdot)$ is the softmax function. Furthermore, layer normalization and residual connection are applied to enhance the robustness and generalization of the model \cite{he2016deep}. More specifically, the multi-modal feature of the item after self-attention is represented as
\begin{align}
\mathbf{i}_{m}^{hs}=ly\_norm  (\mathbf{i}_{m}^{hom}+\mathbf{i}_{m}^{h-self}),
\end{align}
where $ly\_norm(\cdot)$ denotes the layer normalization function.

The cross-attention module is designed to explore the inter-modal influences, and the cross-attention multi-modal features differ from various viewpoints. Therefore, the item's cross-attention visual features and textual features are calculated separately as
\begin{align}
\mathbf{i}_{v}^{h-cross}=softmax(\frac{ (\mathbf{i}^{hs}_{t}\mathbf{W}^{Q}_{t} )^{\top }(\mathbf{i}^{hs}_{v}\mathbf{W}^{K}_{v} )}
{\sqrt{d} } )\mathbf{i}^{hs}_{v}\mathbf{W}^{V}_{v},\\
\mathbf{i}_{t}^{h-cross}=softmax(\frac{ (\mathbf{i}^{hs}_{v}\mathbf{W}^{Q}_{v} )^{\top }(\mathbf{i}^{hs}_{t}\mathbf{W}^{K}_{t} )}
{\sqrt{d} } )\mathbf{i}^{hs}_{t}\mathbf{W}^{V}_{t}.
\end{align}
As in the self-attention post-processing step, residual connection and layer normalization are employed. Formally,
\begin{align}
\mathbf{i}_{m}^{hsc}=ly\_norm  (\mathbf{i}_{m}^{hs}+\mathbf{i}_{m}^{h-cross}).
\end{align}
Note that we use the dropout mechanism in all attention modules to enhance the expressiveness of the model based on the experience of~ \cite{vaswani2017attention}. Since users' multi-modal preferences are not directly obtained, we do not fine-tune them, as evidenced by the experiments.

\subsection{Heterography Relation Learning}
Consistent with previous work \cite{hu2024modality, guo2024lgmrec, wei2019mmgcn}, we conduct graph convolution operations by leveraging  LightGCN~\cite{he2020lightgcn}. Furthermore, the same interaction graph structure $\mathcal{G}_{ui}$ could be employed by different multi-modal and ID information for $\mathcal{G}_{ui}^m$ and $\mathcal{G}_{ui}^{id}$ individually, which operates to ensure non-interference with each other. 

More specifically, the user and item multi-modal features at layer $l+1$ are represented in $\mathcal{G}_{ui}^m$ separately as
\begin{align}
\mathbf{u}_{m}^{( l+1 ) }  =\sum_{i \in \mathcal{N}_{u} }
\lambda_{n} \mathbf{i}_{m}^{(l)},
& \quad
\mathbf{i}_{m} ^{( l+1 ) } =\sum_{u \in \mathcal{N}_{i} }
\lambda_{n} \mathbf{u}_{m}^{(l)},
\end{align}
where $\mathbf{u}_{m}^{( 0) } =\mathbf{u}_{m}^{hom}$, $\mathbf{i}_{m}^{( 0) } =\mathbf{i}_{m}^{hsc}$,  $\mathcal{N}_{u}$ and $\mathcal{N}_{i}$ denote the set of neighbors of the user $u$ and the item $i$ in $\mathcal{G}_{ui}^m$, respectively. The symmetric normalization factor $\lambda_{n} = \frac{1}{\sqrt{|\mathcal{N}_{u}| } \sqrt{|\mathcal{N}_{i}| }}$  is employed to prevent the problem of gradient vanishing or explosion as the number of layers of the graph convolution layer stacks up ~\cite{kipf2016semi}. By combining the neighbor information aggregated across all layers, we obtain the multi-modal representation of the user and the item,
\begin{align}
\bar{\mathbf{u}} _{m}  =\frac{1}{L+1} \sum_{l=0}^{L}\mathbf{u}_{m}^{(l)},& \quad
\bar{\mathbf{i}}_{m}  =\frac{1}{L+1} \sum_{l=0}^{L}\mathbf{i}_{m}^{(l)},
\end{align}
where $L$  denotes the number of layers of LightGCN. It is the same way that we obtain the representation of the user ID $\bar{\mathbf{u}} _{id}$ and the item ID $\bar{\mathbf{i}} _{id}$ after message passing in $\mathcal{G}_{ui}^{id}$ by leveraging GNNs.

\subsection{Refined Contrastive Learning}
Through multi-modal contrastive learning, not only could the self-supervised signals be introduced to alleviate the data sparsity, but the consistency of modalities in the user-item interaction mode could also be enhanced \cite{jiang2024diffmm, guo2024lgmrec}. However, simply contrasting modalities causes the loss of recommendation-relevant information in modal-unique features, and may introduce noise in modal-shared features~\cite{liang2024factorized,dufumier2024align,wang2022rethinking}, thereby impairing recommendation performance. To resolve the above concerns, we introduce meta-network and orthogonal constraint techniques after contrastive learning to further tackle the modal-shared and modal-unique features.

\subsubsection{Multi-modal Contrastive Learning}
Consistent with previous work \cite{jiang2024diffmm, guo2024lgmrec}, we maximize the mutual information between the two modal feature views with the help of InfoNCE \cite{oord2018representation}. Formally, the item multi-modal contrastive learning loss is defined as
\begin{align}
 \mathcal L^{i} _{cl} =\sum_{i\in \mathcal I } -\log\frac{\exp(sim(\bar{\mathbf{i}}_{v}, \bar{\mathbf{i}}_{t})/\tau ) }
{\sum_{i^{'}\in \mathcal I } \exp(sim(\bar{\mathbf{i}}_{v}, \bar{\mathbf{i}}'_t)/\tau )} ,
\end{align}
where $sim(\cdot ,\cdot)$ is the cosine similarity function, and $\tau$ is the temperature coefficient. Similarly, we can define the contrastive learning loss on the user side $\mathcal L^{u} _{cl}$. At last, the total multi-modal contrastive learning loss could be obtained as $\mathcal L_{cl}=\mathcal L^{i} _{cl}+\mathcal L^{u} _{cl}$.

\subsubsection{Modal-shared Meta-network} Simply utilizing features after multi-modal contrasts ignores noise in modal-shared features, such as the “girl” in Figure \ref{fig: intro}, further harming recommendation accuracy. To filter out the noise in modal-shared and derive recommendation-relevant features, we design a meta-network to extract valuable information in modal-shared features, which is inspired by \cite{chen2023heterogeneous, zhu2022personalized}.

We first splice the contrasting multi-modal features and align the dimensions to obtain the modal-shared meta-knowledge. Formally, the meta-knowledge modal-shared feature of the item is denoted as
\begin{align}
\mathbf{i}_{share}=\mathbf{W}^i_{share}(\bar{\mathbf{i}}_{v}||\bar{\mathbf{i}}_{t})+\mathbf{b}^i_{share},
\end{align}
where $\mathbf{W}^i_{share}\in \mathbb{R}^{d\times 2d}$, $\mathbf{b}^i_{share}\in \mathbb{R}^{d}$, and the notation $||$ denotes the concatenation operation. Next, we employ a meta-network to extract recommendation-relevant knowledge and parametrically preserve knowledge into customized transformation matrices. Our proposed meta-neural network could be represented as
\begin{align}
\mathbf{W}^{i_1}_{share}=g^{i_1}(\mathbf{i}_{share}),
& \quad
\mathbf{W}^{i_2}_{share}=g^{i_2}(\mathbf{i}_{share}),
\end{align}
 where $g^{i_1}(\cdot), g^{i_2}(\cdot)$ are meta knowledge learners that comprise a two-layer feed-forward neural network with PReLU activation function. And, $\mathbf{W}^{i_1}_{share} \in \mathbb{R}^{d\times k},\mathbf{W}^{i_2}_{share} \in \mathbb{R}^{k\times d}
$ are customized transformation matrices where modal-shared features are extracted knowledge through the meta-network to filter out noise. The hyperparameter $k$ is the rank of the transformation matrices with restriction to $k<d$, which effectively controls the trainable parameters while enhancing the robustness of the model. To further combine the extracted knowledge by the meta-network with the item representation, inspired by the bridge function \cite{chen2023heterogeneous, zhu2022personalized}, we migrate them to the item ID information. Formally, the transferred modal-shared feature of the item $\bar{\mathbf{i}}_{share}$ is expressed as
\begin{align}
\bar{\mathbf{i}}_{share}=\mathbf{W}^{i_1}_{share} \mathbf{W}^{i_2}_{share} \bar{\mathbf{i}} _{id}.
\end{align}
Since item ID is learned after homography and heterography relationships, we believe it may contain semantic and behavioral information. Hence, we sum it $\bar{\mathbf{i}}_{id}$ with the obtained modal-shared features $\bar{\mathbf{i}}_{share}$ to get the item modal-shared representation $\mathbf{i}^f_{share}$,
\begin{align}
\mathbf{i}^f_{share}=\delta(\bar{\mathbf{i}}_{share})+\bar{\mathbf{i}}_{id}.
\end{align}
where $\delta(\cdot)$ is the PReLU activation function, enhancing feature expressiveness \cite{he2015delving}. Similarly, we get the user's meta-knowledge modal-shared preference $\mathbf{u}_{share}$ and its transferred counterpart $\bar{\mathbf{u}}_{share}$. By fusing ID, we obtain the final user modal-shared representation $\mathbf{u}^f_{share}$.
Note that the complexity of this component is on par with that of the vanilla GNN due to the typically small $k$~\cite{chen2023heterogeneous}.

\subsubsection{Modal-unique Orthogonal Constraint}
As shown in Figure \ref{fig: intro},  “the fashionable twist design” in the image, the combined utilization of different modal-unique and recommendation-relevant information further highlights the complementary nature of multi-modal.

Considering orthogonal constraint loss achieves non-overlapping information between features by encouraging individual features to retain unique information \cite{liu2017adversarial}. Inspired by \cite{bousmalis2016domain, hazarika2020misa,zheng2024mulan}, we impose an orthogonal constraint between multi-modal features to distinguish modal-unique information. Formally, the item multi-modal orthogonal constraint loss between modalities could be denoted as
 \begin{align}
\mathcal L^i_{ort}=\sum_{i\in \mathcal I }\left \| \bar{\mathbf{i}}_{v}^{\top}  \bar{\mathbf{i}}_{t} \right \| ^{2}_F ,
\end{align}
where $\left \| \cdot  \right \| ^{2}_F$ is the squared Frobenius norm. 
We represent item visual and textual features after orthogonal constraints with $\bar{\mathbf{i}}_{v-uni}$ and $\bar{\mathbf{i}}_{t-uni}$, respectively. The same operation is taken to compute the orthogonal loss of the user $\mathcal L^u_{ort}$. At last, we sum to get the total multi-modal orthogonal constraint loss $\mathcal L_{ort}=\mathcal L^u_{ort}+\mathcal L^i_{ort}$.

After obtaining the modal-unique information, we optimize it by utilizing Bayesian Personalized Ranking (BPR)~\cite{rendle2009bpr} to retain the recommendation-relevant information. We align with the treatment of transferred modal-shared features by summing the ID with the modal-unique information. The fusion process is represented as
\begin{align}
\mathbf{i}^f_{m-uni}=\delta(\bar{\mathbf{i}}_{m-uni})+\bar{\mathbf{i}}_{id}.
\end{align}
where $m \in \{v,t\}$, and $\delta(\cdot)$ denotes the PReLU activation function. Then, we splice each modal-unique feature to get the final item modal-unique feature  $\mathbf{i}^f_{uni}=\mathbf{i}^f_{v-uni}\|\mathbf{i}^f_{t-uni}$. Similarly, we can obtain the final user modal-unique preference $\mathbf{u}^f_{uni}$.

\subsection{Prediction and Optimization} Finally, we splice modal-shared and modal-unique features separately to form the final representation, as
\begin{align}
\mathbf{u}^\ast =\mathbf{u}^f_{share}\|\mathbf{u}^f_{uni},
& \quad
\mathbf{i}^\ast =\mathbf{i}^f_{share}\|\mathbf{i}^f_{uni}.
\end{align}
We predict the likelihood of interaction between the user $u$ and the item $i$ by calculating the value of inner product, i.e., $\hat{y}_{ui}  = {\mathbf{u}}^{\ast\top } \mathbf{i}^{\ast}$.

BPR is employed as our principal loss, popularly utilized in recommendation tasks~\cite{wei2019mmgcn, guo2024lgmrec}. And the BPR loss enables REARM to better retain more recommendation-relevant information in multi-modal-unique features. Specifically, we construct a triplet $(u, i, i')$ with a user $u$ and two items, where one item $i$ is observed to have an interaction with the user $u$  and the other item $i'$ is not, 
\begin{align}
\mathcal{R}=\{(u,i,i')|u\in \mathcal{U}, i\in \mathcal{I},i'\notin\mathcal{I}\}.
\end{align}
Formally, the BPR loss function could be expressed as
\begin{align}
\mathcal{L}_{bpr} = \sum_{(u,i,i') \in \mathcal{R}}- ln \psi (\hat{y}_{ui} - \hat{y}_{ui'}), 
\end{align}
where  $\psi(\cdot)$ is the sigmoid function. For better optimization of the model, combining the multi-modal contrastive loss and the orthogonal constraint loss, our final loss function is denoted as
\begin{align}
\mathcal{L} = \mathcal{L}_{bpr}+\lambda_{cl}\mathcal{L}_{cl}+\lambda_{ort}\mathcal{L}_{ort}+\lambda_p \| \Theta \|^2_2, 
\end{align}
where $\lambda_{cl}$ and $\lambda_{ort}$ are hyperparameters that control contrastive and orthogonal loss, respectively. $\Theta$ denotes all model parameters, and hyperparameter $\lambda_p$ controls the impact of $L2$ regularization.

\section{Evaluation}

\begin{table}[tp]
  \centering
  \setlength{\tabcolsep}{3mm}
   \renewcommand\arraystretch{1}
  \caption{Statistics of the three datasets.}
   % \vspace{-10pt}
    \begin{tabular}{ccccc}
    \toprule
    Dataset & \#User & \#Item & \#Interaction & Sparsity \\
    \midrule
    Baby  & 19,445 & 7,050 & 160,792 & 99.88\% \\
    Sports & 35,598 & 18,357 & 296,337 & 99.95\% \\
    Clothing & 39,387 & 23,033 & 278,677 & 99.97\% \\
    \bottomrule
    \end{tabular}%
  \label{Statistics}%
  \vspace{-10pt}
\end{table}%

\subsection{Experimental Settings}
\subsubsection{Evaluation Datasets} Consistent with previous studies \cite{ zhou2023tale, yu2023multi, hu2024modality, guo2024lgmrec}, we conduct experiments on three publicly available and widely used Amazon datasets: Baby, Sports, and Clothing. We filter users and items by utilizing a 5-core as the threshold for each dataset. These three datasets provide rich multi-modal feature data (visual modality and textual modality). In this work, we use the 4,096-dimensional visual features and 384-dimensional textual features that have been preprocessed and released in prior work~\cite{zhou2023mmrec}. The statistics for the three datasets are presented in Table \ref{Statistics}.

\subsubsection{Baseline Methods} To evaluate the effectiveness of REARM, we compare it with state-of-the-art models. Based on whether or not multi-modal utilized, they could be categorized into the general recommendation models (BPR~\cite{rendle2009bpr}, LightGCN ~\cite{he2020lightgcn}, ApeGNN~\cite{zhang2023apegnn}, and MGDN~\cite{hu2024mgdcf}), and multi-modal recommendation models (VBPR~\cite{he2016vbpr}, MMGCN~\cite{wei2019mmgcn}, DualGNN~\cite{wang2021dualgnn}, GRCN~\cite{wei2020graph}, LATTICE~ \cite{zhang2021mining}, BM3~\cite{zhou2023bootstrap}, SLMRec~\cite{tao2022self}, MICRO~\cite{zhang2022latent}, MGCN~\cite{yu2023multi}, DiffMM~\cite{jiang2024diffmm}, FREEDOM~\cite{zhou2023tale}, LGMRec~\cite{guo2024lgmrec}, DRAGON~\cite{zhou2023enhancing}, and MIG-GT~\cite{hu2024modality}).

\begin{table*}[ht]
\centering
	\setlength{\belowcaptionskip}{0cm}
  \setlength{\tabcolsep}{6pt}
\caption{Top-$k$ recommendation performance comparison of different models. The best and the second-best performances are marked with boldface and underlining, respectively.}
\begin{tabular}{ccccccccccccc}
    \toprule
    Dataset & \multicolumn{4}{c}{Baby}      & \multicolumn{4}{c}{Sports}    & \multicolumn{4}{c}{Clothing} \\
    \midrule
    Metric & \multicolumn{1}{l}{R@10} & R@20 & N@10 & N@20 & R@10 & R@20 & N@10 & N@20 & R@10 & R@20 & N@10 & N@20 \\
    \midrule
BPR      & 0.0357          & 0.0575          & 0.0192          & 0.0249          & 0.0432          & 0.0653          & 0.0241          & 0.0298          & 0.0206          & 0.0303          & 0.0114          & 0.0138          \\
LightGCN & 0.0479          & 0.0754          & 0.0257          & 0.0328          & 0.0569          & 0.0864          & 0.0311          & 0.0387          & 0.0361          & 0.0544          & 0.0197          & 0.0243          \\
ApeGNN   & 0.0501          & 0.0775          & 0.0267          & 0.0338          & 0.0608          & 0.0892          & 0.0333          & 0.0407          & 0.0378          & 0.0538          & 0.0204          & 0.0244          \\
MGDN     & 0.0495          & 0.0783          & 0.0272          & 0.0346          & 0.0614          & 0.0932          & 0.0340          & 0.0422          & 0.0362          & 0.0551          & 0.0199          & 0.0247          \\
    \midrule
VBPR     & 0.0423          & 0.0663          & 0.0223          & 0.0284          & 0.0558          & 0.0856          & 0.0307          & 0.0384          & 0.0281          & 0.0415          & 0.0158          & 0.0192          \\
MMGCN    & 0.0421          & 0.0660          & 0.0220          & 0.0282          & 0.0401          & 0.0636          & 0.0209          & 0.0270          & 0.0227          & 0.0361          & 0.0154          & 0.0154          \\
DualGNN  & 0.0513          & 0.0803          & 0.0278          & 0.0352          & 0.0588          & 0.0899          & 0.0324          & 0.0404          & 0.0452          & 0.0675          & 0.0242          & 0.0298          \\
GRCN     & 0.0532          & 0.0824          & 0.0282          & 0.0358          & 0.0599          & 0.0919          & 0.0330          & 0.0413          & 0.0421          & 0.0657          & 0.0224          & 0.0284          \\
LATTICE  & 0.0547          & 0.0850          & 0.0292          & 0.0370          & 0.0620          & 0.0953          & 0.0335          & 0.0421          & 0.0492          & 0.0733          & 0.0268          & 0.0330          \\
BM3      & 0.0564          & 0.0883          & 0.0301          & 0.0383          & 0.0656          & 0.0980          & 0.0355          & 0.0438          & 0.0422          & 0.0621          & 0.0231          & 0.0281          \\
SLMRec   & 0.0521          & 0.0772          & 0.0289          & 0.0354          & 0.0663          & 0.0990          & 0.0365          & 0.0450          & 0.0442          & 0.0659          & 0.0241          & 0.0296          \\
MICRO    & 0.0584          & 0.0929          & 0.0318          & 0.0407          & 0.0679          & 0.1050          & 0.0367          & 0.0463          & 0.0521          & 0.0772          & 0.0283          & 0.0347          \\
MGCN     & 0.0620          & 0.0964          & 0.0339          & 0.0427          & 0.0729          & 0.1106          & 0.0397          & 0.0496          & 0.0641          & 0.0945          & 0.0347          & 0.0428          \\
DiffMM & 0.0623          & 0.0975          & 0.0328          & 0.0411          & 0.0671          & 0.1017          & 0.0377          & 0.0458          & 0.0522         & 0.0791          & 0.0288          & 0.0354          \\ 
FREEDOM  & 0.0627          & 0.0992          & 0.0330          & 0.0424          & 0.0717          & 0.1089          & 0.0385          & 0.0481          & 0.0629          & 0.0941          & 0.0341          & 0.0420          \\
LGMRec   & 0.0644          & 0.1002          & 0.0349          & 0.0440          & 0.0720          & 0.1068          & 0.0390          & 0.0480          & 0.0555          & 0.0828          & 0.0302          & 0.0371          \\
DRAGON   & 0.0662          &\underline { 0.1021 }         & 0.0345          & 0.0435          & 0.0752          &\underline { 0.1139 }         & 0.0413          & \underline{ 0.0512}    & \underline{ 0.0671}    & \underline { 0.0979}          & \underline{ 0.0365}    & \underline {0.0443}    \\
MIG-GT   & \underline {0.0665}          &\underline { 0.1021}          & \underline {0.0361}          & \underline {0.0452}          &\underline { 0.0753}          & 0.1130          & \underline {0.0414}    & 0.0511          & 0.0636          & 0.0934          & 0.0347          & 0.0422          \\
    \midrule
\rowcolor{gray!15} REARM    & \textbf{0.0705} & \textbf{0.1105} & \textbf{0.0377} & \textbf{0.0479} & \textbf{0.0836} & \textbf{0.1231} & \textbf{0.0455} & \textbf{0.0553} & \textbf{0.0700} & \textbf{0.0998} & \textbf{0.0377} & \textbf{0.0454} \\
\bottomrule
\end{tabular}
\label{tab: total}
\end{table*}

\subsubsection{Evaluation Protocols} To obtain fair results for the experimental evaluation, we adopt the same settings as in previous studies~\cite{hu2024modality, zhou2023enhancing, zhou2023tale}. Specifically, we assess recommendation performance with the two most common evaluation metrics, Recall and Normalized Discounted Cumulative Gain (NDCG). Regarding the dataset partition, we randomly split the history interactions according to $8:1:1$ to obtain the train, validate, and test sets. Finally, we evaluate the top-$K$ recommendation with the all-ranking protocol, and report the average performance over all users in the test set for $K = 10$ and $K = 20$. For simplicity, the evaluation metrics can be represented separately as R@K and N@K, where $K\in\{10, 20\}$.

\subsubsection{Implementation Details} To be fair, we set both the user and item embedding dimensions to 64 and the training batch size to 2,048 to optimize all models. We search for optimal parameter ranges in the validation set as reported by their original baseline models for all models, and report the corresponding test set results. To ensure convergence, early stopping and the total epochs are set to 20 and 2,000, respectively. In line with previous work~\cite{yu2023multi, hu2024modality, guo2024lgmrec}, we employ R@20 as the training stop indicator on the validation set. For our model, we simply adjust the number of LightGCN layers from 1 to 10, the user and item modal importance coefficients from 0.1 to 1, the user and item co-occurrence coefficients from 0.1 to 1, the transformation matrix rank from 1 to 10, and the dropout rates of all attention modules from 0 to 1.

\subsection{Overall Performance}
The performance comparison results of all models on three datasets are shown in Table \ref{tab: total}. We observe the following findings,
\begin{itemize}[leftmargin=*]
    \item Our model significantly outperforms all baselines (both general and multi-modal recommendation models) on every dataset. We attribute the remarkable improvement to: 1) We refine current multi-modal contrastive learning with meta-network and orthogonal constraint methods to filter out noise from modal-shared features while preserving modal-unique and recommendation-relevant information. 2) We further mine the user interest graph and item co-occurrence graph based on previous homogeneous
graphs, which enable the model to explore their potential structural and semantic relationships in detail with the help of GNNs.
    \item The majority of multi-modal recommendation models (e.g., MIG-GT, DRAGON, LGMRec, and FREEDOM) significantly outperform general recommendation models (e.g., BPR, MGDN, LightGCN, and ApeGNN), which suggests that the multi-modal information can help the models better to understand the items and the users' multi-modal interests. In addition, DRAGON shows better performance by integrating the user co-occurrence graph and the item semantic graph. It reveals that mining homogeneous relationships among users and items is quite effective, based on which we further explore the user interest graph and item co-occurrence graph to refine the homogeneous relationships.
\end{itemize}

\begin{table}[t]
	\setlength{\belowcaptionskip}{0cm}
  \centering
  \setlength{\tabcolsep}{3.5pt}
  \caption{Ablation study on different modules of the REARM.}
   \begin{tabular}{ccccccc}
    \toprule
    Dataset & \multicolumn{2}{c}{Baby} & \multicolumn{2}{c}{Sports} & \multicolumn{2}{c}{Clothing} \\
    \midrule
    Metric & R@20  & N@20  & R@20  & N@20  & R@20  & N@20 \\
    \midrule
    w/o uu & 0.1055  & 0.0458  & 0.1209  & 0.0544  & 0.0984  & 0.0446  \\
     w/o ii & 0.0972  & 0.0420  & 0.1011  & 0.0453  & 0.0676  & 0.0298  \\
      w/o co & 0.1057  & 0.0458  & 0.1158  & 0.0511  & 0.0968  & 0.0434  \\
       w/o sim & 0.0903  & 0.0402  & 0.0982  & 0.0445  & 0.0708  & 0.0325  \\
     w/o hom & 0.0926  & 0.0403  & 0.0975  & 0.0437  & 0.0653  & 0.0290  \\
    \midrule
    w/o meta & 0.1078  & 0.0471  & 0.1214  & 0.0547  & 0.0995  & 0.0451  \\
     w/o ort & 0.1079  & 0.0470  & 0.1201  & 0.0548 & 0.0983  & 0.0447  \\
     w/o ref & 0.1070  & 0.0467  & 0.1190  & 0.0545  & 0.0982  & 0.0446  \\
     \midrule
\rowcolor{gray!15}     REARM & \textbf{0.1105}  & \textbf{0.0479}  & \textbf{0.1231}  & \textbf{0.0553}  & \textbf{0.0998}  & \textbf{0.0454}  \\
    \bottomrule
    \end{tabular}%
  \label{tab: ablation}%
  \vspace{-10pt}
\end{table}%

\begin{figure*}[t]
 \vspace{-15pt}
    \setlength{\abovecaptionskip}{0cm}
    \setlength{\belowcaptionskip}{0cm}
    \centering  
    \subfigure{ 
            \includegraphics[width=0.23\textwidth]{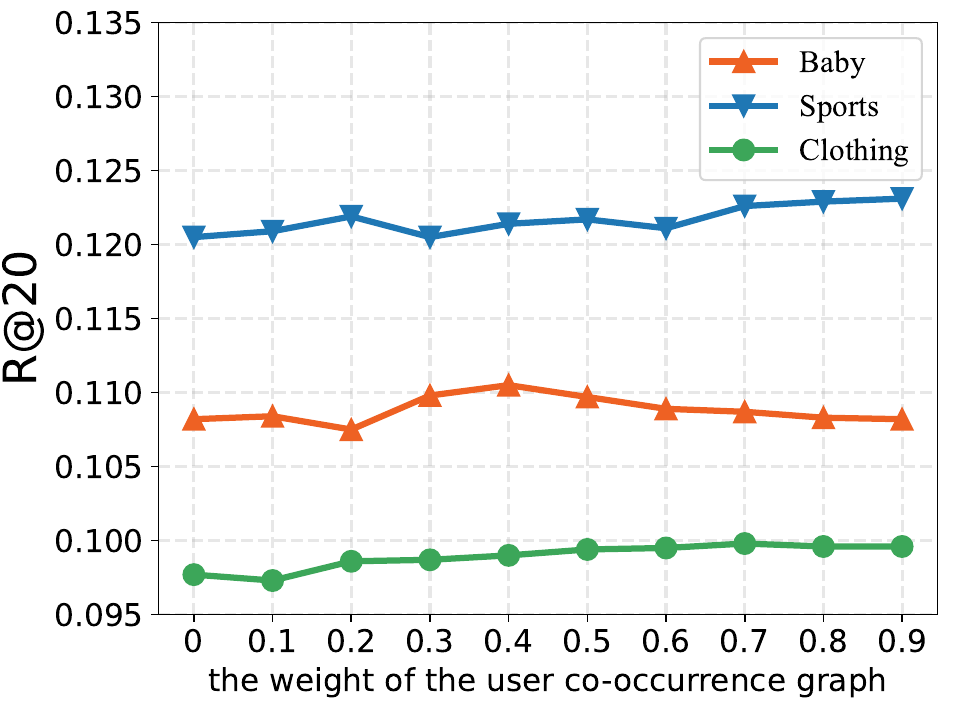}
    }
    \subfigure{ 
            \includegraphics[width=0.23\textwidth]{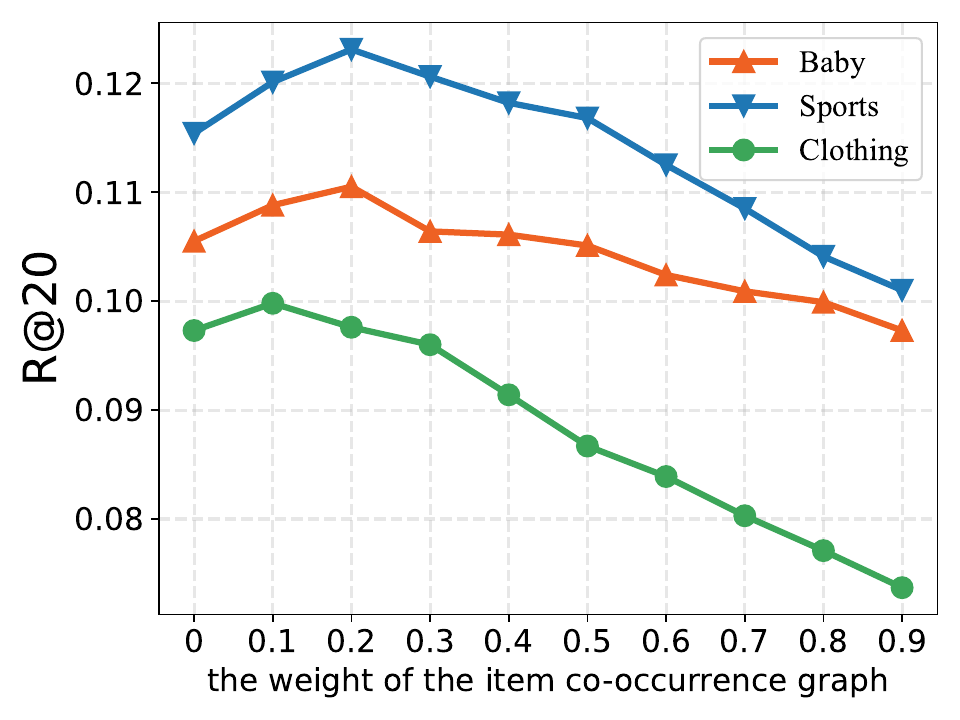}
    }
    \subfigure{ 
            \includegraphics[width=0.23\textwidth]{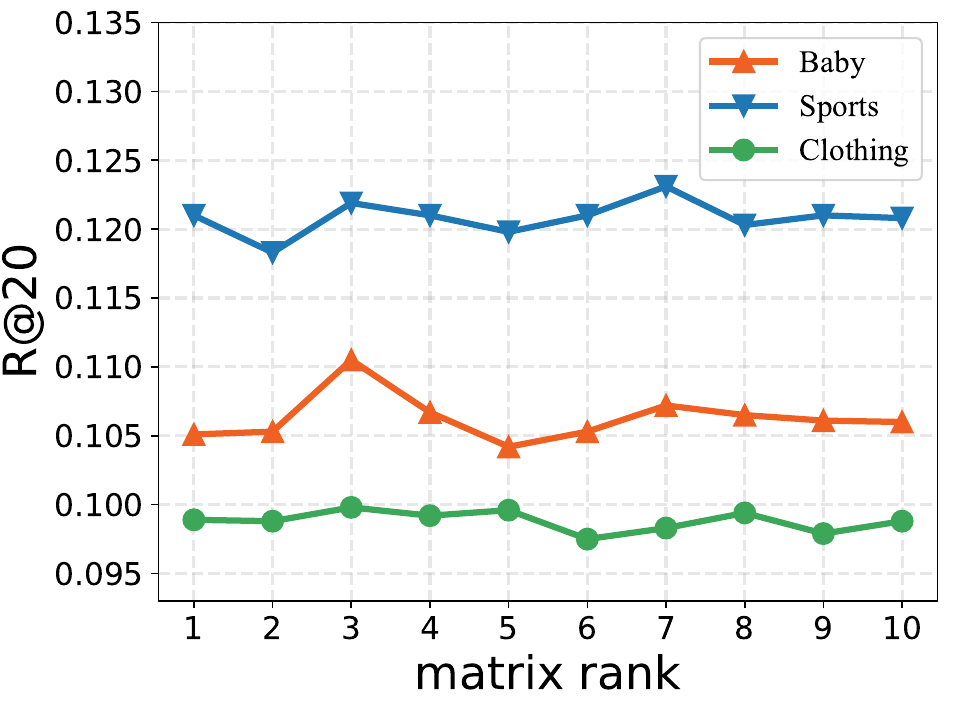}
    }
    \subfigure{ 
            \includegraphics[width=0.23\textwidth]{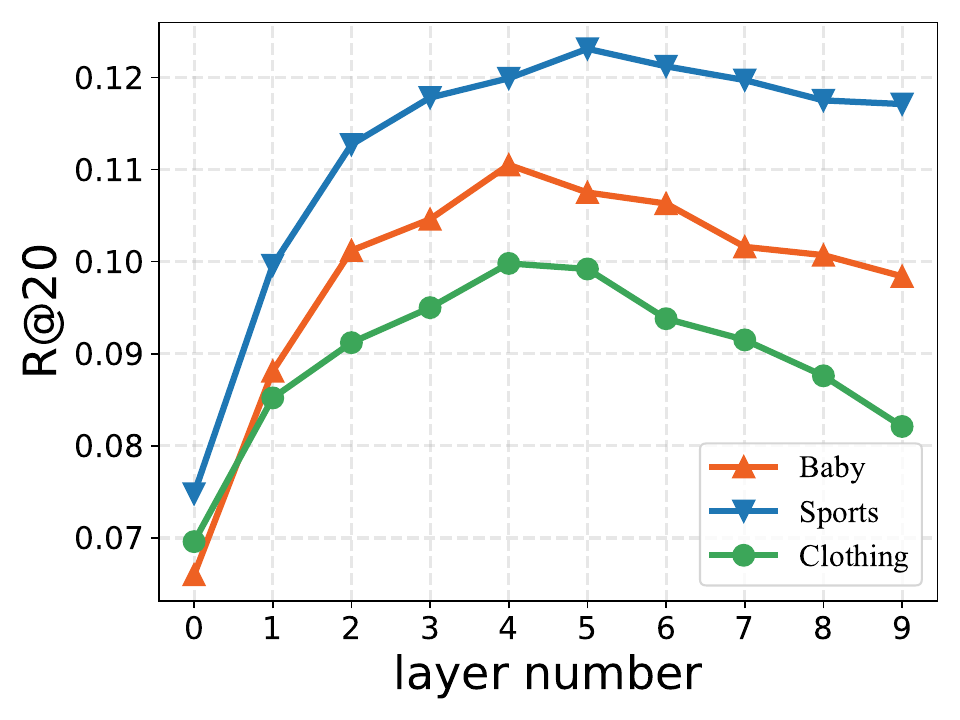}
    }
    \caption{Performance of REARM in terms of R@20 w.r.t the different hyperparameter settings.}   
    \label{fig: hyper} 
    \vspace{-10pt}  
    \Description{...}
\end{figure*}

\begin{figure}[t]
    \setlength{\abovecaptionskip}{0cm}
    \setlength{\belowcaptionskip}{0cm}
    \centering  
    \subfigure{ 
        \centering    
        \includegraphics[width=0.47\columnwidth]{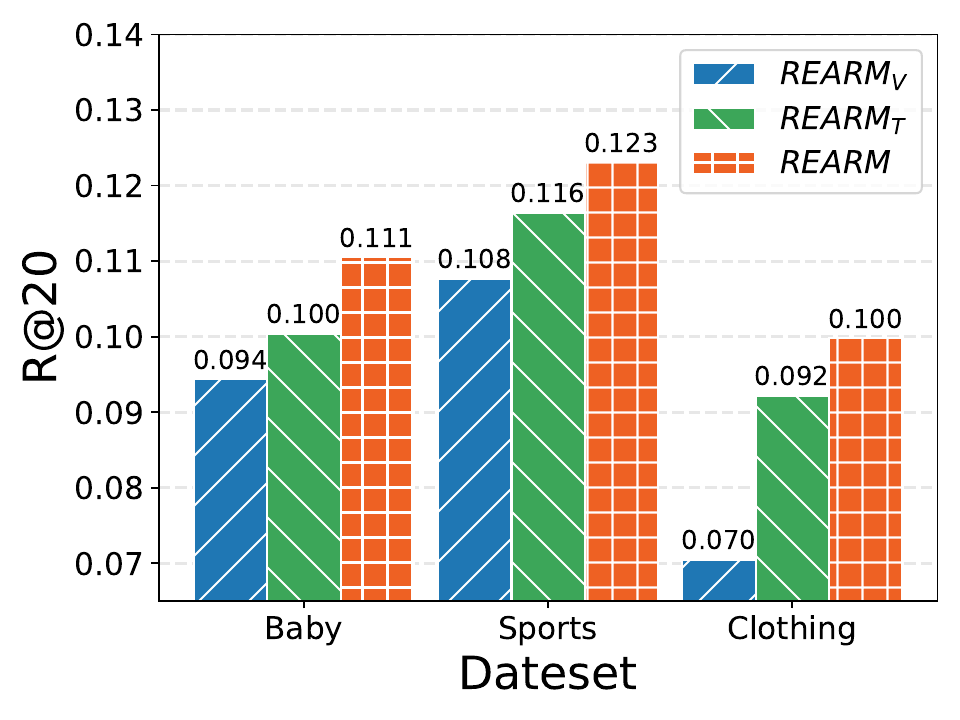}      
    }
    \subfigure{   
        \centering    
        \includegraphics[width=0.47\columnwidth]{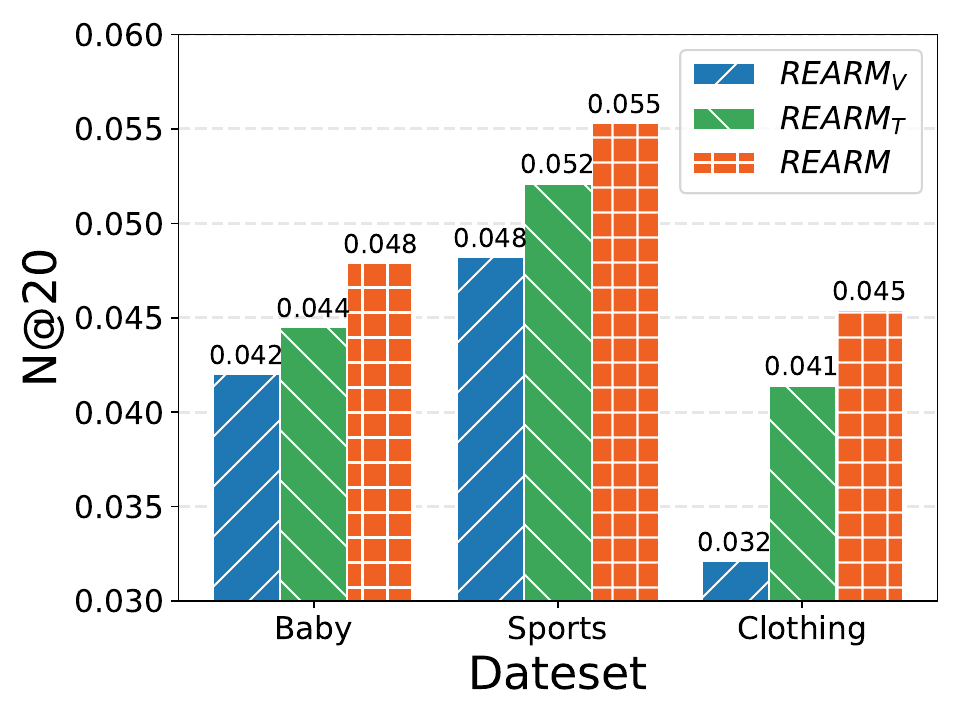}
    }
    \caption{Comparison of different feature modalities.}   
    \label{fig: model_vt} 
    \vspace{-10pt}  
    \Description{...}
\end{figure}

\subsection{Ablation Study}
\subsubsection{Effect of Different Components of REARM}
To analyze the impact of various homographs on the model performance, we categorize them into: no user homograph (w/o uu), no item homograph (w/o ii), no co-occurrence homographs (w/o co), no similarity homographs (w/o sim), and no homographs at all (w/o hom). For the refinement of multi-modal contrastive learning, we split it into: no meta-networks (w/o meta), no orthogonal constraints (w/o ort), and no refinement (w/o ref).

\begin{itemize}[leftmargin=*]
\item The variant w/o hom achieves the worst results on both datasets, which indicates that the introduction of homographs can greatly improve performance. Furthermore, w/o uu is better than w/o ii, and w/o co is better than w/o sim. This illustrates the greater importance of item homogeneous graphs and semantic relations.
 \item  For refining the contrastive learning module, removing either the meta-learning or the orthogonal constraint module results in performance degradation. This further points to the importance of filtering modal-shared feature noise and focusing on recommendation-related information in modal-unique features.
\end{itemize}

\subsubsection{Effect of Different Modalities of REARM}
To study the contribution of diverse modalities to our proposed model, we conduct separate experiments for visual and textual modalities, denoted as $REARM_V$ and $REARM_T$, respectively. The experiment results are shown in Figure \ref{fig: model_vt}, and we could make the following findings.
\begin{itemize}[leftmargin=*]
\item Neither $REARM_V$ nor $REARM_T$ achieves the desired results under both metrics, which verifies that each modality is critical since both of them contain different and unique information.

\item $REARM_T$  is better than $REARM_V$ in all datasets, especially in the Clothing. This is reasonable, as texts contain more refined information and less noise compared to images, which could help users to better select clothes with satisfactory sizes and styles.
\end{itemize}

\subsection{Hyperparameter Analysis}
\subsubsection{Effect of the Weight of the User Co-occurrence Graph} We search for the proportion of the user co-occurrence graph in the final user homograph in $\{0,0.1,0.2,0.3,0.4,0.5,0.6,0.7,0.8,0.9\}$. As shown in Figure \ref{fig: hyper}, we find that compared to the user co-occurrence graph, the user interest graph with a certain weight could better improve the model effect, which is consistent with our motivation.

\subsubsection{Effect of the Weight of the Item Co-occurrence Graph} We explore the contribution of the item co-occurrence graph by searching in $\{0,0.1,0.2,0.3,0.4,0.5,0.6,0.7,0.8,0.9\}$. The optimal weight values for various datasets are different, as seen in Figure \ref{fig: hyper}, which is expected and demonstrates that mining inter-item co-occurrence relationships could enrich the representation of users and items.

\subsubsection{Effect of Matrix Rank in Modal-shared Meta-network.} We effectively control the number of model parameters and enhance the model robustness by the rank of the matrix, searched within $\{1,2,3,4,5,6,7,8,9,10\}$. From Figure \ref{fig: hyper}, we observe that the rank of the matrix for the Sports dataset is 7, which is the largest. A possible reason is that it has the most interactions in three datasets, which may require more model capacity to capture richer interactions.

\subsubsection{Effect of the Number of Layers in the Interaction Graph.} In the user-item interaction graph, we search for the number of propagation layers from 0 to 9. We find that the optimal number of layers for all datasets is greater than 3, as illustrated in Figure \ref{fig: hyper}, which is inconsistent with previous studies, where most of them are at 2 layers. We speculate that it is related to the attention mechanism, which retains valuable information by exploring intra- and inter-modal relationships, thereby allowing information to be propagated to a greater extent via higher-order GNNs.

\begin{figure}[t]
	\centering
	\includegraphics[width=\columnwidth]{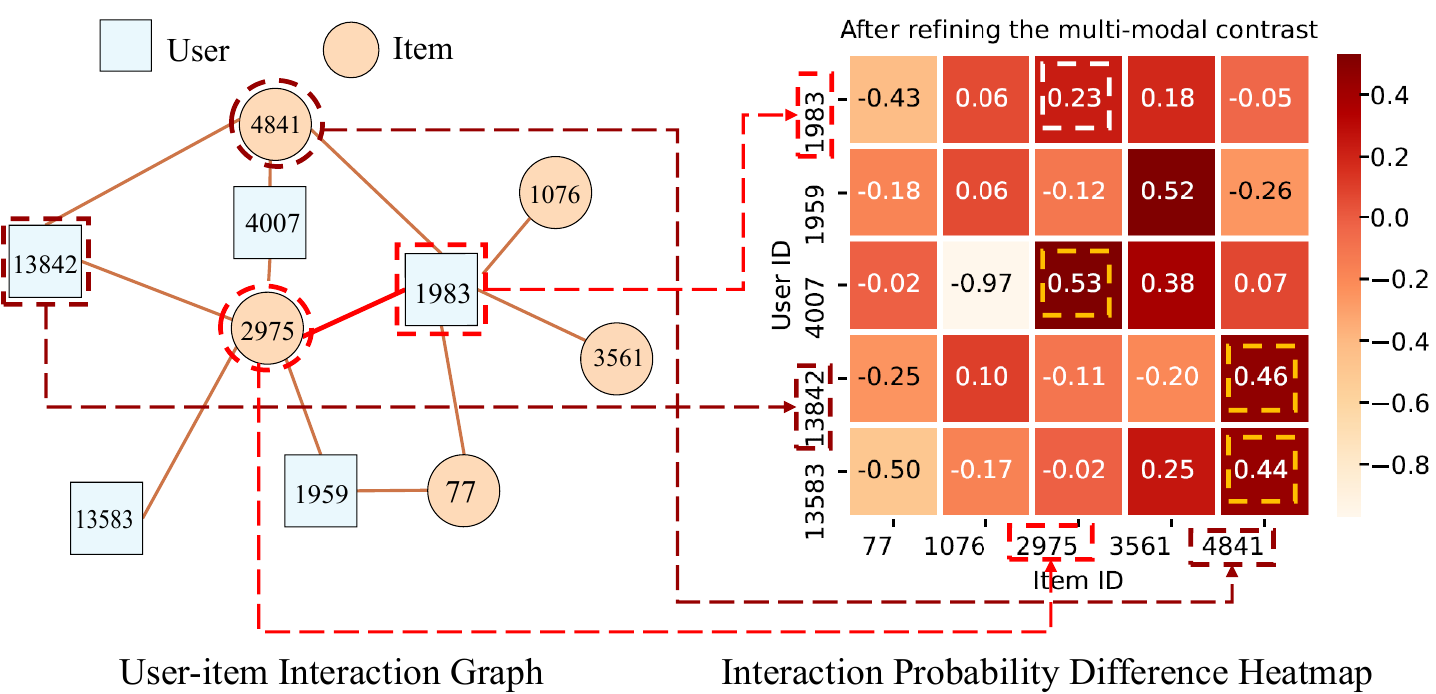}
 \caption{Difference heatmap of interaction probabilities before and after refining multi-modal contrastive learning.}
\vspace{-15pt}
\label{fig: visual}
\Description{...}
\end{figure}

\subsection{Visualization Analysis}

For further investigation of the importance of refining the multi-modal contrastive learning,  we select a portion of users and items from the Baby dataset and calculate the probability of interactions before and after refining contrastive learning, respectively. By differentiating, we get a heatmap of the interaction probability difference.

As shown in Figure \ref{fig: visual}, the deeper the color of the heatmap, the more it indicates that REARM predicts a higher probability of interactions. The yellow boxes in the heatmap indicate that they have been observed to interact in the training set, e.g., user $u_{4007}$ and item $i_{2975}$. Compared to the existing methods, REARM considers the noise in modal-shared and modal-unique and recommended information, rendering it a better predictor. For example, REARM further predicts the high likelihood of interaction between user $u_{1983}$ and item $i_{2975}$, which is confirmed to exist in the test set.

\section{Conclusion}
In this paper, we propose a novel framework for \textbf{R}\textbf{E}fining multi-mod\textbf{A}l cont\textbf{R}astive learning and ho\textbf{M}ography relations  (\textbf{REARM}). A meta-network is designed to denoise the modal-shared features, and the orthogonal constraint loss is utilized to retain the modal-unique and recommendation-relevant information to refine the current multi-modal contrastive learning methods. We also explore potential structural and semantic relationships in user interest and item co-occurrence graphs to enrich representations based on existing homogeneous graphs. Extensive evaluation on three real-world datasets shows that REARM outperforms state-of-the-art baselines.

%%
%% The next two lines define the bibliography style to be used, and
%% the bibliography file.
\bibliographystyle{ACM-Reference-Format}
\balance
\bibliography{sample-base}

\end{document}